\documentclass[aps,twocolumn,showpacs,preprintnumbers,amsmath,amssymb]{revtex4}

\usepackage{dcolumn}
\usepackage{bm}

\newcommand{\Tr}{\text{\,Tr\,}}
\newcommand{\TeV}{\text{\,TeV}}

\def\cut{\Lambda}
\def\Ucut{\Lambda_{\rm U}^{~}}

\def\C{{\mathcal C}}
\def\N{{\mathcal N}}
\def\T{{\mathcal T}}

\newcommand{\beq}{\begin{equation}}
\newcommand{\eeq}{\end{equation}}
\newcommand{\be}{\begin{equation}}
\newcommand{\ee}{\end{equation}}
\newcommand{\beqa}{\begin{eqnarray}}
\newcommand{\eeqa}{\end{eqnarray}}
\newcommand{\ba}{\begin{array}}
\newcommand{\ea}{\end{array}}

\def\leqq{\leqslant}
\def\geqq{\geqslant}        
\def\RE{\Re\mathfrak{e}}

\def\dif{\partial }
\def\to{\rightarrow}
\def\To{\Rightarrow}

\def\dis{\displaystyle}
\def\f{\frac}

\def\[{\left[}
\def\]{\right]}
\def\({\left(}
\def\){\right)}
\def\lg{\left\lgroup}
\def\rg{\right\rgroup}

\def\G{{\mathcal G}}
\def\H{{\mathcal H}}

\begin{document}

\preprint{\large hep-ph/0311177}


\title{Unitarity of Little Higgs Models Signals New Physics of UV Completion}


\author{\sc Spencer Chang\,\,}
\email{chang@physics.harvard.edu}
\affiliation{Jefferson Physical Laboratory, 
             Harvard University, Cambridge, MA  02140, USA}
\author{\sc Hong-Jian He\,\,}
\email{hjhe@physics.utexas.edu}
\affiliation{Center for Particle Physics, 
             University of Texas at Austin, Austin, TX 78712, USA}




\begin{abstract}
\noindent
The ``Little Higgs''  opens up a new avenue 
for natural electroweak symmetry breaking
in which the standard model Higgs particle is realized 
as a pseudo-Goldstone boson and thus is generically light.
The symmetry breaking structure of the Little Higgs models 
predicts a large multiplet of (pseudo-) Goldstone bosons 
and their low energy interactions below the 
ultraviolet (UV) completion scale
\,$\cut \sim 4\pi f \sim O(10)\,\TeV$, 
where \,$f$\, is the Goldstone decay constant.
We study unitarity of the Little Higgs models by systematically analyzing
the high energy scatterings of these (pseudo-)Goldstone bosons. We reveal
that the collective effect of the Goldstone scatterings 
via coupled channel analysis tends to
push the unitarity violation scale $\Ucut$ significantly below the
conventional UV scale \,$\cut \sim 4\pi f$\, as estimated by
naive dimensional analysis (NDA). 
Specifically, \,$\Ucut \sim (3-4)f$,\, lying in
the multi-TeV range for \,$f\sim 1$\,TeV.
We interpret this as an encouraging sign that the upcoming LHC 
may explore aspects of Little Higgs UV completions, and we 
discuss some potential signatures.  
The meanings of the two estimated UV scales
$\Ucut$ (from unitarity violation) and $\cut$ (from NDA) 
together with their implications for an effective field theory analysis
of the Little Higgs models are also discussed.
\end{abstract}

\pacs{11.30.Ly, 12.60.Cn, 12.60.Fr \hfill 
      Preprint$^\#$: HUTP-03/A075, UTHEP-03-19}     
%


\maketitle


\noindent  
{\bf 1. Introduction}
\vspace*{3mm}

The Standard Model (SM) with an elementary Higgs scalar 
is a remarkably simple theory, but despite the simplicity, 
it still successfully accommodates all known experimental
data (aside from neutrino oscillations).  
However, the hierarchy problem\,\cite{Hi} puts the
naturalness and completeness of this theory in doubt.  
Already at one-loop level, quadratic
radiative corrections to the Higgs mass parameter 
destabilize the weak scale, pulling it up to the 
intrinsic ultraviolet (UV) cutoff.  
At best, the SM is an effective field theory behaving naturally only
up to an UV cutoff $\Lambda_{\rm SM}$ 
that could be higher than the weak scale by merely a loop factor, 
\,$\Lambda_{\rm SM} \sim 4\pi v \simeq 3$\,TeV.

This hierarchy problem (or naturalness problem) has motivated most 
of the major extensions of the SM since the seventies. 
The two earliest and best known directions are 
dynamical symmetry breaking\,\cite{DSB} and
the addition of supersymmetry\,\cite{SUSY}.  More recently,
theories with large or small extra dimensions\,\cite{EXD} 
have been used to eliminate the hierarchy problem.
These avenues are quite rich and have been explored in depth.

The newest addition to this list of candidates is an attractive idea called
the ``Little Higgs'' \cite{Arkani-Hamed:2001nc,Arkani-Hamed:2002qy, 
Low:2002ws,Kaplan:2003uc,Chang:2003un,Chang:2003zn}.  
Little Higgs theories seek to solve a {\it little hierarchy,} 
by only requiring the Higgs mass 
be safe from one-loop quadratic divergences.  
In this mechanism, the extended global symmetries enable each interaction to
treat the Higgs particle as a Goldstone boson.  
However, once all interactions are turned on, 
the Higgs becomes a pseudo-Goldstone boson\,\cite{HPGB}. Thus quadratic
divergences in the mass parameter can only appear at two-loops and higher.   
This allows the theory to be natural with an UV cutoff up to two-loop factors 
above the weak scale,
roughly $\Lambda \sim (4\pi)^2 v \sim 10-30$\,TeV.   The required
particle content and interactions are usually quite economical;  there may
be new heavy gauge bosons ($W'$, $Z'$ and $B'$ for instance), new heavy quarks 
($t'$ and possible exotics), and new heavy scalars (electroweak singlets, 
triplets and/or extended Higgs doublet sector).

Many Little Higgs models have been constructed, most of which take
just the minimal solution towards stabilizing the little hierarchy.  
This approach requires a very minimal addition of extra particles and interactions.
At first glance, both experimentalists and theorists might find this
approach depressing, since this just predicts a sparsely filled little 
desert at the LHC.
However, as we will show in this Letter, the situation luckily seems much better.  
In fact, a new scale in the multi-TeV range is found to demand new physics 
beyond that required by the minimal Little Higgs mechanism.

To begin, we can take inspiration from our knowledge of the SM.  
After observing the $W$ and $Z$ gauge bosons, we could wonder
whether their mere existence predicts any new physics to be discovered.
The lesson here is well known.  Since the scattering amplitudes for longitudinal 
weak bosons grow with energy, perturbative unitarity would be 
violated at a critical energy  
$E=\Ucut$ in the absence of Higgs 
boson\,\cite{Smith-73,DM-73,Cornwall-74,LQT-77,Veltman-77,Chanowitz:1985hj}.    
The classic unitarity analysis determines this energy scale as
$\Ucut \simeq 
 1.2$\,TeV\,\cite{DM-73,LQT-77,Veltman-77,Chanowitz:1985hj,Donoghue,Willenbrock}.  
Note that this is noticeably lower than the cutoff scale for strong dynamics,
\,$\cut \sim 4\pi v \simeq 3$\,TeV, as estimated by 
naive dimensional analysis (NDA)\,\cite{NDA,GDA}.

The possible resolutions to this unitarity crisis are well known.  
If a Higgs scalar exists, the Higgs-contributions to the scattering amplitude  
cut off the growth in energy.  Alternatively, 
if strong dynamics breaks the electroweak symmetry, possible 
new vector particles (such as techni-$\rho$\,'s) will save unitarity.  
Imposing perturbative unitarity, these new states 
must appear below or around the scale $\Ucut\simeq 1.2$\,TeV 
for the high energy theory to make sense.
Independent of details in the UV completion, 
this bound ensures new physics to be seen at LHC energies.

Essentially the same lesson can be relearnt for the Little Higgs models.  
The low energy dynamics of the Little Higgs theories are described by 
the leading Lagrangian under the momentum expansion, which is the analog
of the two-derivative operator in the usual chiral Lagrangian.  
Due to the two derivatives,
the scattering amplitude of these scalars is expected to grow as $E^2$,
and will eventually violate unitarity at an energy $E=\Ucut$.  So far,  
the only difference from the SM case is the symmetry breaking structure.  
The different effective chiral Lagrangians will predict different
interaction strengths and relations which determine the unitarity bound.
Most importantly, the bound $\Ucut$  
points to the UV completion scale of the Little Higgs mechanism, 
and in analogy with the SM, is expected to be at accessible 
energy scales, lower than the NDA cutoff 
\,$\Lambda \sim 4\pi f  \sim 10$\,TeV.  
Moreover, because the breaking of extended global symmetries of the 
Little Higgs models results in a large number of additional 
(pseudo-)Goldstones in the TeV range, 
we expect the collective effects of the Goldstone boson scatterings
in a coupled channel analysis
to further push down the unitarity bound $\Ucut$.

The rest of this Letter is organized as follows. We first perform a generic
unitarity analysis for a class of Little Higgs models in Sec.\,2, 
and then carry out an explicit unitarity study for the Littlest Higgs
model of $SU(5)/SO(5)$ in Sec.\,3.  We discuss the potential
new physics signals in Sec.\,4, which  
is not intended to be exhaustive, but just gives a flavor of the 
possible phenomenology at the LHC.  This section ends 
with a discussion of the interpretation and implications for the unitarity
violation scale versus the NDA cutoff scale.
Finally, we conclude in Sec.\,5.

\vspace*{5mm}
\noindent  
{\bf 2. Unitarity of Little Higgs Models: A Generic Analysis}
\vspace*{3mm}

As described in the introduction, Little Higgs models 
predict new physics in the TeV range, such as new gauge bosons 
and new fermions.  However, there can be substantial variation 
in these extra ingredients and thus their analysis is usually   
model dependent.  On the other hand, the 
symmetry breaking structure of a given Little Higgs theory is 
completely determined.  
For instance, the scalars 
in the Littlest Higgs model\,\cite{Arkani-Hamed:2002qy} 
arise from the global symmetry breaking  \,$SU(5)\to SO(5)$.\,  
This guarantees the existence of 14 ``light''
(pseudo-)Goldstone bosons, most of which are expected in the TeV range. 
At leading order in the momentum expansion,
the interactions of these Goldstones are completely 
fixed by the global symmetry breaking pattern.
This allows us to perform a generic analysis of 
the Goldstone boson scatterings and the corresponding unitarity bounds.
Note that the local symmetries (as well as the fermion sector) 
in the Little Higgs theories can vary, but according 
to the power counting\,\cite{Weinberg,power} they do not affect our
analysis of the leading Goldstone scattering amplitudes. 
So we can apply our generic unitarity formula to each given theory 
and derive the predictions.

The setup is rather simple.  As mentioned above, a Little Higgs model
is defined by breaking its global symmetry $\G$ down to a subgroup $\H$.
This guarantees the existence of \,$|\G|-|\H| = \N$\, Goldstone bosons, 
denoted by $\pi^a$  ($a = 1,\, \cdots$,\,${\mathcal N}$).  
At the lowest order of the derivative expansion\,\cite{Weinberg}, 
the Goldstone interactions are fully fixed by the
symmetry breaking structure,
\begin{eqnarray}
\mathcal{L}_{\text{KE}}= 
\dis\frac{\,\,f^2\,}{8} \Tr \left| \partial_\mu \Sigma \right|^2. 
\label{eq:kinetic}
\end{eqnarray} 
In this expression, we define the nonlinear field
$\,\Sigma \equiv \exp\[2i\pi^a T^a/f\]$,\, 
where \,$\Tr(T^aT^b) = \delta^{ab}$\,
ensures the canonical normalization for the 
$\pi^a$'s. The specific form of the broken generators $T^a$ depends 
on the particular model under consideration.
The scale $f$ is the Goldstone decay constant  
and is usually taken to be order $0.7-1$\,TeV\, for naturalness.  
Note that the factor of $1/8$ is a consequence of the normalization
$\Tr(T^aT^b) = \delta^{ab}$ and the definition for $\Sigma$.
Changing the factor $1/8$ will correspond to a simple rescaling of $f$.    
We note that in general the \,$\partial_\mu$\,'s 
should be raised to covariant 
derivatives by gauge invariance.   However, 
since we will be concerned only with the leading Goldstone scatterings
(instead of the more involved gauge boson scatterings), it is
enough to include the partial derivatives.  
This restriction also does not weaken the analysis  
because power counting\,\cite{power} shows that the leading energy growth of 
the Goldstone scattering amplitudes completely arises from the derivative
terms and is independent of the gauge couplings. 
Finally, we note that the only Little Higgs
models which cannot be described by this Lagrangian are the Simple
Group Little Higgses\,\cite{Kaplan:2003uc}.  
This is due to the fact that in those models,
the vacuum expectation value  
$\left< \Sigma\right>$ is not unitary  
and leads to a different structure.

Expanding Eq.\,(\ref{eq:kinetic}) up to quartic Goldstone interactions, we  
arrive at
\begin{equation}
\label{eq:L-4pi}
\dis\!\mathcal{L}_{\text{KE}}
=\frac{1}{2} \partial_\mu \pi^a \partial^\mu \pi^a 
+\frac{\,\Gamma^{abcd}}{3f^2}
(\partial^\mu \pi^a) \pi^b (\partial_\mu \pi^c) \pi^d + O(\pi^5)
\end{equation}
where we have defined
\begin{equation}
\Gamma^{abcd} \,\equiv\, \Tr\left[ T^a T^b T^c T^d -  T^a T^c T^b T^d \right].
\end{equation}
To proceed with a coupled channel analysis, we will consider a 
canonically normalized singlet state under $\H$, consisting of
$\N$ pairs
of Goldstone bosons,
\begin{equation}
\label{eq:S-state}
\dis
\left| S\right> ~=~  
\sum_{a=1}^{\N} \frac{1}{\sqrt{2\N\,}}
\left| \pi^a \pi^a \right> \,,
\end{equation} 
where the factor \,$1/\sqrt{2}$\, is conventionally used to account for  
the identical particle states.
The state $\left| S\right>$ is a singlet since the $\pi^a$\,'s 
form a real representation of the $\H$ symmetry in non-Simple Group models.
Since the $\pi^a$\,'s also form an irreducible 
representation of $\H$,  
this is the only singlet formed from two $\pi^a$\,'s.
The scattering amplitude $\,\T\[S\to S\]\,$ will contain $\N^2$ number of
individual $\pi\pi\to\pi\pi$ channels, and is expected to be the
largest amplitude for deriving the optimal unitarity bound.
For instance, experience with the QCD $SU(2)$ chiral Lagrangian
or the SM Higgs sector shows that  
the isospin singlet channel of \,$\pi\pi$\, scattering results in
the strongest unitarity 
bound\,\cite{LQT-77,Chanowitz:1985hj,Donoghue,Willenbrock}.  
We also note that among the $\pi^a$\,'s there are would-be Goldstone bosons
whose scattering describes the corresponding
scattering of the longitudinal gauge bosons [such as $(W_L,\,Z_L)$ and
$(W_L',\,Z_L',\,B_L')$] in the high energy range ($s \gg m_W^2,m_{W'}^2$)
via the equivalence theorem\,\cite{Cornwall-74,LQT-77,Chanowitz:1985hj,ET}.  
So, at high energies our analysis is equivalent to a 
unitary gauge analysis.

Using the interaction Lagrangian in Eq.\,(\ref{eq:L-4pi}), we can readily determine
the singlet scattering amplitude at tree level,
\begin{equation}
\label{eq:T-SS}
\dis\T\[ S \to S \] = \frac{\C}{\,\N f^2\,}\; s \,,
\end{equation}
where we have defined the group-dependent coefficient
\begin{equation}
\label{eq:C}
\C ~~=~ \dis\sum_{a,b=1}^{\N}\Gamma^{aabb} ~.
\end{equation}
To derive this result, we have used the relation for Mandelstam
variables \,$s+t+u \approx 0$\, 
after ignoring the small pion masses relative to the large energy 
scale $\sqrt{s}$\,.  Here we note that because  
\,$\Gamma^{aaaa}  = 0$,\,  only the $\N (\N-1)$ inelastic
channels, \,$\pi^a\pi^a\to \pi^b\pi^b$\, ($a\neq b$), contribute.

It is now straightforward to compute the 0th partial wave amplitude 
from Eq.\,(\ref{eq:T-SS}), 
\begin{equation}
\!\! a_0^{~}\[ S \to S\] 
 = \dis\frac{1}{32\pi}\!\int_{-1}^1\!\! dz \; P_0(z)\T(s,z) 
\dis = \frac{\C}{\,16\pi\N f^2\,}\,s \,,
\end{equation}
which, as expected, grows quadratically with the energy and  
is subject to the unitarity constraint,
\begin{equation}
\label{eq:UC}
\dis |\RE \,a_0^{~}| ~< ~ \f{1}{2}\,.
\end{equation}
Hence, we find
that perturbative unitarity holds for energy scales 
\begin{equation}
\dis
\sqrt{s} \,~<~\, \sqrt{\frac{\,8\pi\N \,}{| \C |}} \,f 
~\equiv~ \Ucut \,.
\end{equation} 
Since $\C$ tends to scale as $\N^{3/2}$ for large $\N$, 
the unitarity bound should scale as $\N^{-1/4}$ \cite{Sekhar}.  
Hence, we expect the unitarity bound to be 
quite low since $\N$ is reasonably large in the Little Higgs models.

Using this general formula, we can readily compute 
the coefficient $\C$ and determine
the unitarity bounds on the various Little Higgs theories.  
We compile our results in Table\,\ref{table:unitarity}.
Note that for moose models, there is a 
four times replicated non-linear sigma model structure. 
But, we have chosen to analyze only  
one of the non-linear sigma model fields.  
Any interaction between the different non-linear
sigma model fields is model-dependent, 
so this restriction is consistent with our approach.

Table\,\ref{table:unitarity}  shows that
indeed the Little Higgs models generically contain a large
number of Goldstone bosons, ${\mathcal N}=O(10-20)$,\,  
and our unitarity bound $\Ucut$ is significantly lower than
the conventional cutoff of the theory, 
\,$\Lambda \sim 4\pi f \simeq 12.6 f$,\,  as estimated by NDA. 
The observation that the unitarity violation scale 
turns out much lower than $\cut$
is an encouraging sign, indicating that aspects of the Little Higgs  
UV completions may be possibly explored at the LHC.  
We will discuss more about the interpretations of our results 
and highlight the possible collider signatures in Sec.\,4.

\begin{table*}
\caption{Summary of unitarity bounds in various Little Higgs theories.
\label{table:unitarity}}
\begin{center}
\begin{tabular*}{.85\textwidth}{@{\extracolsep{\fill}}l||c|c|c|c|c|c|c}
\hline
\hline
& & & & & & &\\[-2mm]
~~~Little Higgs Model & G~  & H~ & $\N$~~ & $|{\cal C}|$~~ 
                      & $\Ucut/f$~ & $m_{W'}/f$~ & $m_{t'}/f~~~$ 
\\[1.5mm]
\hline
\hline
& & & & & & &\\[-2.4mm]
~~~Minimal Moose \cite{Arkani-Hamed:2001nc}     & 
  $SU(3)^2$ & $SU(3)$~ & ~8~~ & 24~~ & 2.89~    & 2.37~   & 1~~~ \\[1.5mm]
\hline
& & & & & & &\\[-2.4mm]
~~~Littlest Higgs \cite{Arkani-Hamed:2002qy}    & 
  $SU(5)$~ & $SO(5)$~ & 14~~ & 35~~ & 3.17~     & 1.67~    & 2~~~ \\[1.5mm]
\hline
& & & & & & &\\[-2.4mm]
~~~Antisymmetric Condensate \cite{Low:2002ws}~  & 
  $SU(6)$~ & $Sp(6)$~ & 14~~ & 26~~ & 3.68~     & 1.67~    & 2~~~ \\[1.5mm] 
\hline
& & & & & & &\\[-2.4mm]
~~~$SO(5)$ Moose \cite{Chang:2003un}            & 
  $SO(5)^2$ & $SO(5)$~ & 10~~ & 15~~ & 4.09~    & 3.35~    & $\sqrt{2}$~\,~~~ \\[1.5mm] 
\hline
& & & & & & &\\[-2.4mm]
~~~$SO(9)$ Littlest Higgs \cite{Chang:2003zn}& 
  $SO(9)$~ & $SO(5)\otimes SO(4)$~ & 20~~ & 35~~ & 
  3.79~    & 2.37~                 & 2~~~ \\ [1.5mm]
\hline
\hline
\end{tabular*}
\end{center}
\vspace*{-5mm}
\end{table*}

To add a reference frame for the unitarity bounds in 
Table\,\ref{table:unitarity},   we also give
the masses of the $W'$ gauge boson and the $t'$ quark 
(using our current normalization of $f$).  
For the gauge boson, 
the mixing angle between the two $SU(2)$ gauge couplings 
has been set to \,$\theta = 1/5$.\,  
To scale to a different angle $\theta_{\rm new}$,
just multiply by 
\,$\sin{(2/5)}/ \sin{2\theta_{\rm new}}$.\,  
A relatively small mixing angle is required since electroweak
precision analysis restricts 
\,$m_{W'} \gtrsim 1.8 \TeV$\,\cite{Chang:2003un,GSW03}.  
For the $t'$ quark, we have minimized its mass, corresponding to 
maximizing the naturalness; in the particular case of   
two Higgs doublet models we have set \,$\sin\!\beta = 1$\,  
(for other $\beta$ values, just divide by $\sin\!{\beta}$\,).

A striking feature of Table\,\ref{table:unitarity}\,  
is that   \,$2m_{W'} > \Ucut$\, holds for
almost all Little Higgs models except the 
Antisymmetric Condensate model\,\cite{Low:2002ws}  
where $\Ucut$ is only slightly higher 
than the corresponding value of $2m_{W'}$.
Such a low $\Ucut$ means that for the center of mass
energy \,$\sqrt{s}<\Ucut$,\, 
the $W'W'$ scattering processes will not be kinematically allowed.
From the physical viewpoint, 
this strongly suggests that additional new particles 
(having similar mass range) have to co-exist with $W'$\,'s in the same
effective theory so that their presence can properly restore the unitarity. 
But these new states should enter the Little Higgs theory in such a way 
as to ensure the cancellation of one-loop quadratic
divergences\,\cite{sphjh2}. 
From the technical viewpoint, this obviously implies 
the equivalence theorem 
no longer holds for predicting the $W'_LW'_L$ scattering amplitude by
that of the corresponding Goldstone scattering.
But the exact $W'_LW'_L$ 
scattering amplitude could only differ
from the Goldstone amplitude by $m_{W'}^2/s = O(1)$ terms at most, 
and thus are not expected to significantly affect our conclusion.

\vspace*{5mm}
\noindent  
{\bf 3. Unitarity of the Littlest Higgs Model: An Explicit Analysis}
\vspace*{3mm}

In this section we will explicitly analyze the Littlest Higgs model of 
$SU(5)/SO(5)$ \cite{Arkani-Hamed:2002qy} by writing all Goldstone fields
in the familiar electroweak eigenbasis of the SM gauge group. 
Then we will extract the leading
Goldstone scattering amplitudes and derive the unitarity bounds, 
in comparison with our generic analysis of Sec.\,2.

As mentioned earlier, 
the Littlest Higgs model has the global symmetry breaking
structure \,$SU(5)\to SO(5)$,\, resulting in 14 Goldstone bosons which
decompose under the SM gauge group $SU(2)_W\otimes U(1)_Y$ as
\beq
{\bf 1}_0 \,\oplus\, {\bf 3}_0 \,\oplus\, {\bf 2}_{\pm 1/2} \,\oplus\, {\bf 3}_{\pm 1} \,.
\eeq
Here the
${\bf 1}_0 \oplus {\bf 3}_0$ denotes a real singlet $\chi^0_y$ and a real triplet
$\chi^{\pm,0}$. They will become the longitudinal components of gauge bosons 
$(B',\,W',\,Z')$ 
when the gauged subgroups $[SU(2)\otimes U(1)]^2$ are Higgsed down to the
diagonal subgroup $G_{\rm SM}$. 
The ${\bf 2}_{\pm 1/2}$ includes a Higgs doublet $H$
and ${\bf 3}_{\pm 1}$ a complex Higgs triplet $\Phi$, defined as
\beq
\dis H^T =\! \left\lgroup \ba{c} \dis\pi^+ \\[4mm]
           \dis\f{v+h^0+i\pi^0}{\sqrt{2}}
    \ea \right\rgroup \!,    ~~~~
   \Phi = \!\left\lgroup \ba{cc}
         \phi^{++}                &~\,  \dis\f{\phi^+}{\sqrt{2}}   \\[4mm]
         \dis\f{\phi^+}{\sqrt{2}} &~\,  \phi^0\!-iv'
         \ea \right\rgroup \!,
\eeq
where the would-be Goldstones $\pi^{\pm,0}$ will be absorbed by the light
gauge bosons $(W^\pm,\,Z^0)$ when electroweak symmetry breaking is triggered
by the Yukawa and gauge interactions via the 
Coleman-Weinberg mechanism\,\cite{CW}.
There will be some small mixings between the scalars in $H$
and $\Phi$ due to the nonzero triplet VEV $v'$,
but the condition $M_\Phi > 0$ requires \cite{Han}
\beq
\dis v' \,<\, \f{v^2}{\,4f\,} ~\ll~ v \,, 
\eeq
so that for the current purpose 
it is enough to expand the tiny ratio \,$v'/v$\, 
and keep only its zeroth order at which  
the two sets of Goldstone bosons do not mix. 
This greatly simplifies our explicit analysis.

Collecting all the 14 Goldstone bosons we can write the nonlinear field
$\Sigma = \exp\[i2\Pi/f\]\Sigma_0$ for the $SU(5)/SO(5)$ model 
where the $5\times5$ Goldstone matrix is given by
\beq
\ba{l}
\dis\Pi= 
\lg  
\ba{ccc}
\dis\f{1}{2}X     &  \dis\f{1}{\sqrt{2}}H^\dag   & \Phi^\dag
\\[4mm]
\dis
\f{1}{\sqrt{2}}H  &  \dis\f{2}{\sqrt{5}}\chi_y^0 
                  &  \dis\f{1}{\sqrt{2}}H^*
\\[4mm]
\Phi              &  \dis\f{1}{\sqrt{2}}H^T     & \dis\f{1}{2}X^*
\ea
\rg ,
\ea
\eeq
and
\beq
\!
X \!=\!\! 
\lg  
\ba{cc}
\!\!\dis\chi^0\!-\!\f{\chi^0_y}{\sqrt{5}}  & \,\dis\sqrt{2}\chi^+ \!  
\\[4mm]
\!\dis\sqrt{2}\chi^-   & \!\dis -\chi^0\!-\!\f{\chi^0_y}{\sqrt{5}}\!\! 
\ea
\rg \!\! , ~~\,
\Sigma_0 \! =\!\! 
\lg 
\ba{ccc} 
 \!&\!   &\! \!\mathbf{1}_{2\times2}\! \\[2mm]
 \!&\! 1 &\!                       \\[2mm]
\!\mathbf{1}_{2\times2}\!\! \!&\!  &\!
\ea 
\rg \!\! .\!\!
\eeq
Similar to Eq.\,(\ref{eq:kinetic}), 
we derive the leading order Goldstone boson Lagrangian
\beq
\label{eq:LH0-L}
\ba{l}
\dis
{\cal L}_{\rm KE}  
=\dis\f{\,\,f^2\,}{8}\,{\rm Tr}\left|\partial_\mu\Sigma\right|^2
\\[4mm]
= \dis \f{1}{2}{\rm Tr}\(\dif^\mu\Pi\)^2 \!+\!
       \f{1}{3f^2} {\rm Tr}\!\[(\Pi\dif^\mu\Pi)^2 \!\!-\!
                          (\dif^\mu\Pi)^2\Pi^2\] \!+\!
       O(\Pi^5),
\ea
\eeq
where the the first dimension-4 operator gives 
the canonically normalized kinetic terms
for all Goldstone fields in $\Pi$, and the second term
gives the quartic Goldstone interactions.

To derive the optimal unitarity limit from the Goldstone
scatterings, we will consider a canonically normalized
$SO(5)$ singlet state consisting of 14 pairs of Goldstone bosons,
\beqa
\label{eq:Sall}
\left| S \right> &= & \!\!
\dis\f{1}{\sqrt{28}} \[~
2\left|\pi^+\pi^-\right> + \left|\pi^0\pi^0\right>
+\left|h^0h^0\right> +2\left|\chi^+\chi^-\right>  \right.
\nonumber
\\[0mm]
&& \dis
~~~~~~~ 
+\left|\chi^0\chi^0\right>
+\left|\chi^0_y\chi^0_y\right> + 2\left|\phi^{++}\phi^{--}\right> 
\nonumber
\\[1.5mm]
&& \dis\left.
~~~~~~\,
+2\left|\phi^+\phi^-\right> + 
  \left|\phi^0_1\phi^0_1\right> +
  \left|\phi^0_2\phi^0_2\right> \,
\] ,
\eeqa
where we have defined \,$\phi^0\equiv \phi_1^0 +i\phi^0_2$\,.\, 
This is essentially a re-expression of our general formula
(\ref{eq:S-state})  with all \,$\N =14$\, Goldstone fields 
in the electroweak eigenbasis. 
But the expanded form of the quartic interactions in
(\ref{eq:LH0-L}) is extremely lengthy in the electroweak eigenbasis, 
making the explicit calculation of the whole amplitude $\T[S\to S]$ tedious.
Before giving a full calculation of $\T[S\to S]$,
we will {\it explicitly} expand Eq.\,(\ref{eq:LH0-L}) and illustrate
the unitarity limits for the two sub-systems  $(\chi^a,\,\chi^0_y)$
and $(\pi^{\pm,0},\,h^0)$.
From Eq.\,(\ref{eq:LH0-L}), we derive the
corresponding interaction Lagrangians 
\beq
\ba{l}
{\cal L}_{\rm int}^{\pi h} = 
\dis\!\f{1}{12f^2} \!\left\{  \,
\[ -\,(2vh + h^2) (\dif_\mu\pi^a \dif^\mu\pi^a)\, -
\right.\right.
\\[3mm]
\left.
\,~~~~~ - (\dif_\mu h)^2 {\pi^a}^2
\!+\! 2(v+h)(\dif_\mu h)(\pi^a\dif^\mu\pi^a) \]
\\[3mm]
\,~~~~~ +\!\[(\dif_\mu\pi^+)^2{\pi^-}^2 \!-\! 
[(\dif_\mu\pi^0)^2\!+\!\dif_\mu\pi^+\dif^\mu\pi^-]\pi^+\pi^- 
\right.
\\[3mm]
\left.\left.
~~~~~+2(\pi^0\dif_\mu\pi^0)(\pi^+\dif^\mu\pi^-)
\!-\!{\pi^0}^2(\dif_\mu\pi^+\dif^\mu\pi^-)
\!+\!{\rm H.c.}\]\right\},
\\[4mm]
{\cal L}_{\rm int}^{\chi} = 
\dis\!\f{1}{6f^2} \!
\left\{(\dif_\mu\chi^+)^2{\chi^-}^2 \!-
[(\dif_\mu\chi^0)^2\!+\!\dif_\mu\chi^+\dif^\mu\chi^-]\chi^+\chi^- 
\right.
\\[3mm]
\left.
~~~~~~~+2(\chi^0\dif_\mu\chi^0)(\chi^+\dif^\mu\chi^-)
\!-\!{\chi^0}^2(\dif_\mu\chi^+\dif^\mu\chi^-)
\!+\!{\rm H.c.}\right\},
\ea
\eeq
where the $U(1)$ Goldstone $\chi^0_y$ does not enter
${\cal L}_{\rm int}^{\chi}$ at this order.  
The Goldstones $(\pi^{\pm,0},\,h^0)$ form 
the SM Higgs doublet $H$ which also
has a renormalizable Coleman-Weinberg potential.
But unlike ${\cal L}_{\rm int}^{\pi h}$, this potential only 
contributes constant terms to the Goldstone 
amplitudes and thus do not threaten the unitarity,
especially when the pseudo-Goldstone Higgs $h^0$ is 
relatively light as favored by the electroweak precision data.

The Lagrangian \,${\cal L}_{\rm int}^{\pi h}$\, describes the
leading derivative interactions of the Higgs
doublet $H$, characterized by the Goldstone decay constant $f$
and originated from the global symmetry breaking $SU(5)\to SO(5)$.  
In analogy with the SM case\,\cite{LQT-77}, we find that 
$(\pi^{\pm,0},\,h^0)$
form an electroweak singlet state
$\left| S_{H}\right> = \f{1}{\sqrt{8}}
\[2\left|\pi^+\pi^-\right>+\left|\pi^0\pi^0\right>
  +\left| h^0 h^0\right>
\].
$  
The corresponding $s$-wave amplitude is
\,$a_0[S_H\to S_H] = (3s/64\pi f^2)$,\, 
where we have dropped small terms suppressed by the extra 
factor \,$(v/f)^2 \ll 1$\,.\,  
Imposing the condition (\ref{eq:UC}), we deduce the unitarity limit
\beq
\label{eq:UB-pi-h}
\dis\sqrt{s} \,~<~\, \Ucut = \sqrt{\f{32\pi}{3}}\,f \,\simeq\, 5.79 f \,,
\eeq
which is lower than the NDA cutoff 
\,$\cut \sim 4\pi f$\, by a factor of $2.2$.
Note that contrary to the scatterings of Goldstone $\pi^a$\,'s
(or $W_L/Z_L$\,'s) in the SM, 
the $\pi\pi$ scatterings in the Littlest Higgs
model grow with energy due to the derivative interactions in
${\cal L}_{\rm int}^{\pi h}$.
Next, we turn to the $(\chi^\pm,\,\chi^0)$ system.
The Lagrangian  ${\cal L}_{\rm int}^{\chi}$ 
for the Goldstone triplet is the same 
as the familiar $SU(2)$ chiral Lagrangian. 
So we define the normalized isospin singlet state 
\,$\left|S_{\chi^a}\right> = \f{1}{\sqrt{6}}
\[2\left|\chi^+\chi^-\right>+\left|\chi^0\chi^0\right>
\]$,\,  
and derive its $s$-partial wave amplitude
$\,a_0^{~}\[S_{\chi^a}\to S_{\chi^a}\] = s/(16\pi f^2)$.\,  
Using the condition (\ref{eq:UC}), we arrive at
\beq
\label{eq:UB-X}
\dis\sqrt{s} \,~<~\, \Ucut = \sqrt{8\pi}\,f \,\simeq\, 5.01 f \,,
\eeq
which is lower than \,$\cut \sim 4\pi f$\, by a factor of $2.5$.
%

After the above explicit illustrations, we will proceed
with a full analysis of this model in the electroweak eigenbasis.
The key observation is that the $SO(5)$ singlet state 
$\left| S\right>$ in Eq.\,(\ref{eq:Sall}) can be decomposed into
$4$ smaller orthonormal states formed from two $\pi^a$'s, 
\beq
\label{eq:S4singlets}
\hspace*{-2mm}
\left|S\right> =\dis
\sqrt{\f{2}{7}} \left| S_H\right> +
\sqrt{\f{3}{14}}\left| S_{\chi^a}\right> +
\f{1}{\sqrt{14}} | S_{\chi_y^0}\rangle +
\sqrt{\f{3}{7}} \left| S_\Phi\right> ,
\eeq
each of which is an {\it electroweak singlet} state, defined as
\beqa
\label{eq:4singlets}
\left| S_H\right> 
\!\! &\equiv&\!\! \dis
\f{1}{\sqrt{8}}\sum_{a=1}^4 \left| \pi^a\pi^a\right> 
\nonumber\\
\!\! &=&\!\! \dis\f{1}{\sqrt{8}}\[2\left|\pi^+\pi^-\right> + \left|\pi^0\pi^0\right>
                   +\left|h^0h^0\right>\] ,
\nonumber\\
\left| S_{\chi^a}\right> 
\!\! &\equiv& \!\! \dis
\f{1}{\sqrt{6}}\sum_{a=5}^7 \left| \pi^a\pi^a\right> 
= \f{1}{\sqrt{6}}\[2\left|\chi^+\chi^-\right> + \left|\chi^0\chi^0\right>\] ,
\nonumber\\
| S_{\chi_y^0}\rangle 
\!\! &\equiv &\!\! \dis\f{1}{\sqrt{2}}\left|\pi^8\pi^8\right>
                        = \f{1}{\sqrt{2}}\left|\chi_y^0\chi_y^0\right>,
\\
\left| S_\Phi\right> 
\!\! &\equiv& \!\! \dis
\f{1}{\sqrt{12}}\sum_{a=9}^{14} \left| \pi^a\pi^a\right> 
\nonumber\\
\!\! &=& \!\!\!
\dis\f{1}{\sqrt{12}}\!\[2\left|\phi^{++}\!\phi^{--}\right> \!+\!
                           2 \left|\phi^{+}\!\phi^{-}\right> \!+\!      
                            \left|\phi_1^0\phi_1^0\right> \!+\!
                            \left|\phi_2^0\phi_2^0\right> \] \!.
\nonumber
\eeqa

Now we will perform a full coupled-channel analysis for the Goldstone scatterings 
among these $4$ electroweak singlet states and prove that the maximal eigenchannel
just corresponds to the amplitude $\T [S\to S]$ in Sec.\,2
with $|S\rangle$ given by Eq.\,(\ref{eq:S4singlets}) 
[equivalently, Eq.\,(\ref{eq:Sall}) or (\ref{eq:S-state})]. 
There are $16$ such individual scattering channels in total.    
Denoting each singlet state in Eq.\,(\ref{eq:4singlets}) as
\,$\left|S_j\right> \equiv\dis 
   \f{1}{\sqrt{2{\mathcal N}_j}}\!\!\sum_{a=a_j^{\min}}^{a_j^{\min}-1+{\mathcal N}_j}
   \!\left|\pi^a\pi^a\right> $\, with $j=H,\chi^a,\chi_y^0,\Phi$,
we can now readily derive any amplitude $\T [S_j\to S_{j'}]$
by using the general formulas (\ref{eq:T-SS})-(\ref{eq:C}),
\beq
\label{eq:SS-jj'}
\ba{rcl}
\T[S_j\to S_{j'}] &\, = \,& \dis
\f{{\mathcal C}_{jj'}}{~\sqrt{\N_j\N_{j'}}f^2~}\,s \,,
\ea
\eeq
where 
\,${\mathcal C}_{jj'}    = \sum_{a=a_j^{\min}}^{a_j^{\min}-1+\N_j}
                           \sum_{c=c_{j'}^{\min}}^{c_{j'}^{\min}-1+\N_{j'}}
                           {\mathcal C}^{aacc}
  $\,
will be explicitly evaluated for $SU(5)/SO(5)$.
So, with all the singlet states $\left|S_j\right>$, we deduce a
$4\times 4$ matrix of the leading $s$-wave amplitudes
\beq
\label{eq:S4x4}
{\cal A}_0 ~=~ \dis
\f{s}{\,16\pi f^2\,}
\lg
\ba{cccc}
\dis\f{3}{4}        &~ \dis\f{\sqrt{3}~}{4}  &~   \dis\f{~5}{\sqrt{2}} &~ \dis\sqrt{\f{3}{8}}
\\[4mm]
\dis\f{\sqrt{3}~}{4} &~ 1                    &~   0                    &~ \dis\f{~1}{\sqrt{2}}
\\[4mm]
\dis\f{~5}{\sqrt{2}} &~ 0                    &~   0                    &~ 0
\\[4mm]
\dis\sqrt{\f{3}{8}} &~ \dis\f{~1}{\sqrt{2}}  &~   0                    &~ \dis\f{3}{2}
\ea
\rg .
\eeq
It has the eigenvalues
\,$a_{0j}=\dis\f{s}{16\pi f^2}\(-1,\,\f{1}{2},\,\f{5}{4},\,\f{5}{2}\)$,\,
where the maximum channel
\,$a_0^{\max}=5s/(32\pi f^2)$\,  corresponds to a normalized 
eigenvector \,$(\sqrt{2/7},\,\sqrt{3/14},$ $\,\sqrt{1/14},\,\sqrt{3/7})$,\,
which in this basis is precisely the singlet state in Eq.\,(\ref{eq:S4singlets})!
Imposing the condition (\ref{eq:UC}), we derive the best unitarity
limit for the Littlest Higgs model,
\beqa
\label{eq:UB-best}
\dis\sqrt{s}  &~ < ~& \Ucut 
    = \dis\sqrt{\f{16\pi}{5}}\,f \,\simeq\, 3.17 f \,,
\eeqa
in perfect agreement with the optimal bound in Table\,I.

With the information in Eq.\,(\ref{eq:S4x4}), we can also analyze the optimal 
unitarity limits for all {\it sub-systems} via partial coupled-channel analysis, 
as summarized below.
\beq
\label{eq:UBsum}
\ba{lclc}
\hline\hline
\\[-1.5mm]
~~{\rm Subsystem}&~ \Ucut   ~~~&~~~ 
{\rm Subsystem}&~ \Ucut~~
\\[1.5mm]
\hline
\\[-1.5mm]
~~\{H\}\!\!:&~           5.79f     ~~~~&~~~~
\{H,\,\chi^a\}\!\!:&~  4.35f~~ 
\\[2mm]
~~\{\chi^a\}\!\!:&~      5.01f     ~~~~&~~~~
\{H,\,\Phi\}\!\!:&~    3.69f~~
\\[2mm]
~~\{\Phi\}\!\!:&~               4.09f     ~~~~&~~~~
\{\chi^a,\,\Phi\}\!\!:&~      3.45f~~ 
\\[2mm]
~~\{H,\chi^a,\chi^0_y\}\!\!:&~  3.71f   ~~~~&~~~~
\{H,\chi^0_y,\Phi\}\!\!:&~    3.45f~~
\\[2mm]
~~\{\chi^a,\chi^0_y,\Phi\}\!\!:&~   3.45f  ~~~~&~~~~
\{H,\chi^a,\Phi\}\!\!:&~          3.27f~~  
\\[1.5mm]
\hline\hline
\ea
\eeq 
It clearly shows that as more states are included into
the coupled channel analysis, the unitarity limit $\Ucut$
becomes increasingly stronger and approaches the best
bound (\ref{eq:UB-best}) in the full coupled-channel analysis.
It also demonstrates the limit $\Ucut$ to be fairly robust since 
omitting a few channels does not significantly alter the result.  
Finally, for the subsystems $\{H\}=\{\pi^{\pm,0},\,h^0\}$
and $\{\chi^a\}$, we see that 
Eq.\,(\ref{eq:UBsum}) nontrivially agrees with 
Eqs.\,(\ref{eq:UB-pi-h})-(\ref{eq:UB-X}) derived  
from explicitly expanding (\ref{eq:LH0-L}).

In summary, taking the Littlest Higgs model as an example,
we have explicitly analyzed the unitarity limits from the 
Goldstone scatterings via both partial and full coupled-channel analyses,
with the Goldstone fields defined in the familiar electroweak eigenbasis.
These limits are summarized in Eqs.\,(\ref{eq:UBsum}) and (\ref{eq:UB-best}).
We find that the best constraint (\ref{eq:UB-best}) indeed comes from the 
{\it full coupled-channel analysis} including all $14$ Goldstone fields in the
$SO(5)$ singlet channel (Eq.\,(\ref{eq:S-state}) or (\ref{eq:S4singlets})), 
in complete agreement with Table\,I  (Sec.\,2).  
We have also systematically analyzed the smaller subsystems where some channels
are absent. Most of the resulting unitarity limits in Eq.\,(\ref{eq:UBsum})
are fairly close to the best limit, so    
Eq.\,(\ref{eq:UB-best}) is relatively robust.

\vspace*{5mm}
\noindent  
{\bf 4. Implications for New Physics Signals}
\vspace*{3mm}

As shown in Sec.\,2-3, the unitarity constraints already indicate that Little 
Higgs theories have an important intermediate scale $\Ucut$, 
which is in the multi-TeV region and below the conventional NDA cutoff 
\,$\Lambda\sim 4\pi f$.\,  
Somewhere below $\Ucut$, new particles should appear in order to unitarize the 
Goldstone scattering of $\pi^a$\,'s.  
In particular, the longitudinal $W_LW_L/Z_LZ_L$ scattering 
(or the corresponding Goldstone scattering $\pi\pi\to \pi\pi,hh$) 
will be measured by experiments. 
This process should start to exhibit resonance behavior at least by the 
scale $\Ucut$, although what actually unitarizes the amplitude depends 
upon the UV completion.  
For the case of the Minimal Moose\,\cite{Arkani-Hamed:2001nc}, 
we can rely on our intuition from the QCD-type dynamics.  If it is dynamical
symmetry breaking that generates the $SU(3)^2 \to SU(3)$ breaking, the 
new states should be the analogous vector meson multiplet, i.e., TeV scale
$(\rho,\,K^*,\, \omega,\,\phi)$ particles.  
On the other hand, we could envision a linear sigma model completion 
(with/without supersymmetry).
As an example, 
there could be a scalar $\Sigma$ that transforms as a $(3,\,\bar{3})$ 
and gets a VEV proportional to the $3\times 3$ unit matrix.   
In this case, we can expect new singlets and heavy octet scalars to 
appear in addition to the octet of Little Higgs bosons.  
If the Little Higgs theory respects T-parity (cf. second reference in 
\cite{Chang:2003un}), these new states would have to be even under this 
parity.  This means they can be singly produced and also have restricted 
decay channels, allowing only an even number of T-odd particles 
in the final state. 
So, selecting a specific UV completion can predict a 
very interesting phenomenology.  This direction will be pursued 
further\,\cite{sphjh2}. 
In order to investigate the phenomenology of these new states, realistic UV 
completions should be searched for.  
For instance, Ref.\,\cite{Nelson:2003aj} provides an interesting dynamical 
UV completion, but more constructions should also be actively sought.

One might also wonder if small mixing angles or coupling constants would
render these new states hard to observe experimentally.  
We clarify this by noting that the approximate global symmetry $\H$  
relates the scattering of the $\H$ singlet to the 
scattering of light longitudinal $W/Z$ bosons 
in the following manner.  Neglecting $\H$ breaking effects,
the general amplitude of $\pi\pi$ scattering is given by
\beq
\T(\pi^a\pi^b \to \pi^c\pi^d) ~\,=~\, \dis\sum_j c^{abcd}_j A_j(s,t,u) \,,
\eeq  
where $j$ is a finite integer, 
$c^{abcd}_j$ is a constant tensor invariant under $\H$, and 
$A_j(s,t,u)$ is a kinematic function depending on the Mandelstam variables.  
The $\H$ singlet amplitude is a specific linear combination 
of the kinematic functions.  
At the lowest order, we have seen that these functions grow with $s$ and this 
specific combination needs to be altered at least by $\Ucut$.  
However, longitudinal $W/Z$ scattering is just another linear 
combination of these kinematic functions.  Thus, at the scale $\Ucut$,
unitarizing only the $\H$ singlet scattering but keeping
the SM-type scattering channels unaffected will require an 
accidental cancellation in the group theory space. 
So, generically any new resonance should be shared among all allowed
individual scattering channels even though an amplitude 
for the SM-type channel alone violates unitarity 
at a relatively higher scale\,\cite{Sekhar}.
At worst, a possibly suppressed coefficient should only arise from 
the projection into the SM-type channel, rather than a small
mixing or coupling (up to $\H$ breaking effects).

The scale $\Ucut$ certainly opens up encouraging possibilities 
at the LHC, not only to test the minimal Little Higgs mechanism, 
but also to start probing possible new signs of its UV completion dynamics. 
We note that the unitarity bound $\Ucut\sim (3-4)f$ puts an {\it upper
limit} on the scale of new states which are going to restore the 
unitarity of the Little Higgs effective theory up to the UV scale 
$\sim\!10$\,TeV or above. So the masses of these new states can be
naturally at anywhere between $\sim\!f$ and $\Ucut$, but their
precise values must depend on the detailed dynamics of a given
UV completion. For instance, QCD-like UV dynamics would
predict the lowest new resonance to be a $\rho$-like vector boson
which is expected to be relatively heavy and close to our upper limit $\Ucut$.
But when the UV dynamics invokes supersymmetry, the lowest new
state that unitarizes the $W_LW_L$ scattering would be scalar-like and 
can be substantially below $\Ucut$, 
say $\sim\!0.5f$ according to the lesson of supersymmetric SM.  
(Note that the classic unitarity 
bound for the Higgsless SM only requires 
$\sqrt{s} < \Ucut =\sqrt{8\pi}v\simeq 5.0v\simeq 1.2$\,TeV\,\cite{DM-73,
LQT-77,Veltman-77,Chanowitz:1985hj,Donoghue,Willenbrock}, but the 
minimal supersymmetric SM unitarizes the $W_LW_L$ scattering
by adding 2-Higgs-doublets with the lightest Higgs boson mass 
$M_h \lesssim 130\,{\rm GeV} \simeq 0.5v$ \cite{SUSY}, which is 
typically a factor $\sim\!10$ below $\Ucut$.)
So, it is legitimate to expect the lightest
new state in the UV completion of Little Higgs models to lie 
anywhere in the range 
$0.5f \lesssim M_{\rm new}^{\min} \leqq \Ucut$, 
though its precise mass value is highly model-dependent.
The natural size for the scale $f$ is $\sim\!1$\,TeV 
\cite{Arkani-Hamed:2001nc}--\cite{Chang:2003zn}. 
The updated precision analyses\,\cite{Chang:2003zn,Peskin,GSW03}
showed that the Little Higgs models are readily consistent with
the current data which constrain  $f\gtrsim 0.5-1$\,TeV at 95\%C.L. 
(depending on details of the parameter space in each given model)
\footnote{E.g., it was shown\,\cite{Peskin} that the early precision bound
in the Littlest Higgs model is essentially relaxed by just gauging
the subgroup $SU(2)\times SU(2)\times U(1)$.}, 
so $f$ is allowed to be around its natural size $\sim\!1$\,TeV. 
Taking $f\sim 1$\,TeV for instance, 
we expect the lightest new state to be around 
$0.5\,{\rm TeV} \lesssim M_{\rm new}^{\min} \lesssim 3-4$\,TeV.
So, if lucky, the LHC may produce the lightest new resonance, 
or if it is too heavy, detect the effect of its resonance-tail
(via higher order model-dependent contributions in the low energy
derivative expansion)\,\cite{Mike-WW}. 
But a quantitative conclusion has to be highly model-dependent.
To be conservative, we warn that the limited 
LHC center-of-mass energy does not guarantee the discovery
for such state, 
especially when $M_{\rm new}^{\min}$ is close to the upper limit 
$\Ucut$. 
Further precision probe may be done at future $e^+e^-$
Linear Colliders, and the proposed CERN CLIC with 
$E_{\rm cm}=3-5$\,TeV and 
${\cal L}=10^{35}{\rm cm}^{-2}{\rm s}^{-1}$\,\cite{CLIC}
is particularly valuable.
The definitive probe of the Little Higgs UV dynamics is expected
at the future VLHC 
($E_{\rm cm}=50-200$\,TeV and 
${\cal L}\gtrsim 10^{35}{\rm cm}^{-2}{\rm s}^{-1}$)\,\cite{VLHC}.
Incorporating the new signatures of UV completion into relevant collider 
analyses will expand upon the existing phenomenological 
studies\,\cite{Han,Peskin,Burdman:2002ns}.

Next, we discuss the meanings of the two estimated UV scales, 
$\Ucut$ and $\cut$, 
and their implications for an effective field theory analysis
in the Little Higgs models.
We note that these UV scales are determined by two different measures 
of perturbativity breakdown.  
Our lowest unitarity limit $\Ucut$ is derived from 
the Goldstone scatterings in the singlet channel via the $s$-partial wave.
(Weaker bounds may be obtained for the non-singlet channels via the
higher order partial waves.)
On the other hand, the NDA estimate of the UV cutoff
is based on the consistency of the chiral perturbation expansion,
i.e., one estimates the coefficient of an operator (counter term)
of dimension-$D$ from its renormalization-group running induced by
one-loop contributions of an operator of dimension-$(D-2)$ and 
so on\,\cite{Weinberg,NDA},  
because the former's size should be at least of
the same order as the latter's one-loop contribution 
(about $O(1)/16\pi^2$ multiplied by an $O(1)$ logarithm) 
barring an accidental cancellation.  So one obtains the
{\it original} NDA result\,\cite{NDA},

%
\beq
\label{eq:NDA}
\ba{l}  
~~~~~~
\vspace*{-20mm}
\\  
\hspace*{-5mm}
\dis\f{\,f^2\,}{\cut^2} ~\gtrsim~ \f{\,O(1)\,}{16\pi^2},\, 
~~~\To~~~ \cut ~\lesssim~ 4\pi f \,,
\vspace*{-14mm}
\ea
\eeq

\noindent
which is a conservative {\it upper bound} on the UV cutoff.
The true cutoff for the effective theory should be 
$\min \(\Ucut,\,\cut\)$.
From low energy QCD, the chiral perturbation theory breaks down 
as the energy reaches the $\rho$-resonance at 
$M_\rho = 0.77$\,GeV which is below but still close
to the upper limit \,$4\pi f\simeq 1.2$\,GeV.\, 
So we know this original NDA
upper bound $4\pi f$ describes the UV scale of the low energy QCD
quite well\,\footnote{The best unitarity limit of the low energy
QCD $\pi\pi$ scattering comes from the $I=0$ isospin-singlet channel, 
$\Ucut \simeq \sqrt{8\pi}f \simeq 0.47$\,GeV. This lies
significantly below the upper limit $4\pi f$ by a factor of $2.5$.
It is interesting to note that in the physical spectrum, 
besides the $\rho$ meson, there are good evidences
for a relatively light and broad $\sigma$ meson in the 
$\sim\!0.5$\,GeV range\,\cite{sigma}
which unitarizes the $I=0$ channel and agrees well with the unitarity limit
$\Ucut\simeq 0.47$\,GeV.  The fact that QCD chiral Lagrangian works quite
well is largely because $\sigma$ is a very broad $I=0$ resonance and hard to
detect\,\cite{sigma}.}.
But, the dynamics of Little Higgs UV completions can of course be 
very different from QCD dynamics (or even supersymmetric). 
In fact, for an underlying gauge interaction
with large color $N_c$ and flavor $N_f$,
a Generalized Dimensional Analysis (GDA)\,\cite{GDA,Sekhar} gives
\beq
\label{eq:GDA}
\cut ~\lesssim~ \dis\min 
     \(\f{a}{\sqrt{N_c}},\,\f{b}{\sqrt{N_f}}\)4\pi f \,,
\eeq
where $a$ and $b$ are constants of order $1$. So we see that as long
as $N_c$ or $N_f$ is much larger than that of QCD, the GDA cutoff
will indeed be lower than the original NDA estimate.
Furthermore, the observation that the unitarity of Goldstone scatterings
indicates a lower UV cutoff for the chiral perturbation was
made in \cite{Sekhar}, where it was shown that for
a symmetry breaking pattern $SU(N)_L\otimes SU(N)_R \to SU(N)_V$
($N\geqq 2$), the $\pi\pi$ scattering in the 
$SU(N)_V$-singlet and spin-0 channel  
would impose a unitarity violation scale
\beq
\cut ~\lesssim~ \dis \f{\,\,4\pi f\,\,}{\sqrt{N\,}\,} \,, 
\eeq
signaling a significantly lower UV scale for new resonance formation 
in comparison with the original NDA estimate.  
This is consistent with our current unitarity analysis 
for the Little Higgs models.

Finally, in an effective field theory analysis of the Little Higgs models, 
which UV cutoff is more relevant for suppressing the higher-dimensional 
operators? The precise answer has to be very model-dependent, 
relying on what type of heavy state(s) is integrated out when
generating a given effective operator. 
Without knowing the true UV dynamics, the original NDA estimate 
$\cut\sim 4\pi f$ could be considered as a conservative analysis
where the UV scale is the highest possible.  So far all the 
electroweak precision analyses\,\cite{Chivukula:2002ww,GSW03,Peskin} 
adopted the NDA estimate of $\cut$.  
But we should keep in mind that the actual UV cutoff $\cut$ 
could be significantly lower, as suggested by $\Ucut$, although $\cut$ 
has to be fixed by the underlying dynamics 
[cf. GDA estimate in Eq.\,(\ref{eq:GDA})].
Hence it will be instructive to take 
the two UV scales $\Ucut$ and $\cut\sim 4\pi f$ 
as guidelines and allow the predictions to vary in between.  
The ultimate determination of the UV scale can only come from  
future experiments.

\vspace*{5mm}
\noindent  
{\bf 5. Conclusions}
\vspace*{3mm}

In this Letter, we systematically studied the unitarity constraints 
in various Little Higgs models using a general formalism in Sec.\,2.
Our analysis of the Goldstone scatterings 
is rather generic and mainly independent 
of the choices of parameters, gauge groups and fermion interactions, etc.  
This is because the leading Goldstone interactions in the derivative expansion 
are completely governed by the structure of global symmetry breaking, 
allowing us to perform a coupled channel analysis for the {\it full} Goldstone
sector in a universal way.  We observed that because the global
symmetry breaking in the Little Higgs theories generically predict a large  
number of (pseudo-)Goldstone bosons, their collective effects via coupled
channel analysis of Goldstone scatterings tend to push the unitarity violation
scale $\Ucut$ significantly below the conventional NDA cutoff 
\,$\cut \sim 4\pi f\simeq 12.6f$.\,   
Specifically, \,$\Ucut \sim (3-4)f$\, (cf.~Table\,I),
which puts an {\it upper limit} on the mass 
of the lightest new state, i.e.,  
\,$M_{\rm new}^{\min} \,\leqq\,
\Ucut\sim\! (3-4)$\,TeV~  for ~$f\sim 1$\,TeV\,.

As a comparison, in Sec.\,3
we took the Littlest Higgs model of $SU(5)/SO(5)$ as an example 
and explicitly analyzed the Goldstone scatterings in their electroweak 
eigenbasis. We performed both partial and full coupled-channel analyses.
We derived various unitarity violation limits for this minimal model
and demonstrated that as more Goldstone states are included into the
coupled channel analysis, the unitarity limit 
\,$\Ucut$\,  becomes increasingly stronger, close to the
best bound [cf. Eqs.\,(\ref{eq:UBsum}) and (\ref{eq:UB-best})].
This concrete analysis shows that the optimal unitarity limits 
in Sec.\,2 are fairly robust.

We stress that these tight unitarity limits strongly suggest the 
encouraging possibility of testing the precursors of the 
Little Higgs UV completion at the upcoming LHC (although no guarantee
is implied). 
A definitive test is expected at the future VLHC\,\cite{VLHC}.
In Sec.\,4 we discussed some implications 
for the UV completions and the related collider signatures. 
Finally, we concluded Sec.\,4 by discussing the meanings of the two estimated
UV cutoff scales $\Ucut$ (from unitarity violation) and $\cut$ (from NDA/GDA).
Deciding which estimate to be more sensible in an effective field theory
analysis of Little Higgs models is unclear before knowing the precise UV 
dynamics. Only future experiments can provide an ultimate, definitive answer.

\vspace*{3.2mm}
\noindent
Note added:  
As this work was being completed, a related preprint\,\cite{Mahajan:2003vr}
appeared which did an explicit unitary-gauge calculation of only 
light $W_L/Z_L$ scattering in the Littlest Higgs model.
Unfortunately its result is incorrect due to, for instance, 
mistaking the upper bound on the Higgs triplet VEV which leads
to erroneously large gauge-Higgs triplet couplings.

\vspace*{3.3mm}
\noindent
{\bf Acknowledgments} \\[1.5mm]
It is our pleasure to thank H. Georgi, M. Luty, M. E. Peskin, 
J. Wacker, and especially N. Arkani-Hamed and R. S. Chivukula
for valuable discussions.  
We also thank the organizer of SUSY-2003   
for the hospitality and the stimulating enviroment 
during June\,5-10, 2003 when this work was initiated.   
SC was supported by an NSF graduate student fellowship,
and HJH by the DOE under grant DE-FG03-93ER40757.


\end{document}